%% file: main.tex
\documentclass[sigconf]{acmart}

\usepackage{enumitem}
\usepackage{multirow}
\usepackage{graphicx}
\usepackage{float}
\usepackage{subcaption}
\usepackage{svg}
\usepackage{xcolor}

\usepackage{algorithm}
\usepackage{algpseudocode}

\AtBeginDocument{%
  \providecommand\BibTeX{{%
    \normalfont B\kern-0.5em{\scshape i\kern-0.25em b}\kern-0.8em\TeX}}}

\copyrightyear{2024}
\acmYear{2024}
\setcopyright{acmlicensed}
\acmConference[KDD '24]{Proceedings of the 30th ACM SIGKDD Conference on Knowledge Discovery and Data Mining}{August 25--29, 2024}{Barcelona, Spain}
\acmBooktitle{Proceedings of the 30th ACM SIGKDD Conference on Knowledge Discovery and Data Mining (KDD '24), August 25--29, 2024, Barcelona, Spain}
\acmDOI{10.1145/3637528.3671841}
\acmISBN{979-8-4007-0490-1/24/08}
\definecolor{sequences}{RGB}{112,48,160}
\definecolor{patterns}{RGB}{0,113,125}

\begin{document}

\title[DR4SR]{Dataset Regeneration for Sequential Recommendation}



\author{Mingjia Yin}
\email{mingjia-yin@mail.ustc.edu.cn}
\orcid{0009-0005-0853-1089}
\affiliation{%
  \institution{University of Science and Technology of China \& State Key Laboratory of Cognitive Intelligence}
  \city{Hefei}
  \country{China}
}

\author{Hao Wang}
\authornote{Corresponding author.}
\email{wanghao3@ustc.edu.cn}
\orcid{0000-0001-9921-2078}
\affiliation{%
  \institution{University of Science and Technology of China \& State Key Laboratory of Cognitive Intelligence}
  \city{Hefei}
  \country{China}
}

\author{Wei Guo}
\email{guowei67@huawei.com}
\orcid{0000-0001-8616-0221}
\author{Yong Liu}
\email{liu.yong6@huawei.com}
\orcid{0000-0001-9031-9696}
\affiliation{%
  \institution{Huawei Singapore Research Center}
  \country{Singapore}
}


\author{Suojuan Zhang}
\email{sjzsj@ustc.edu.cn}
\orcid{0009-0003-2193-2288}
\author{Sirui Zhao}
\email{sirui@mail.ustc.edu.cn}
\orcid{0000-0001-8103-0321}
\affiliation{%
  \institution{University of Science and Technology of China \& State Key Laboratory of Cognitive Intelligence}
  \city{Hefei}
  \country{China}
}


\author{Defu Lian}
\email{liandefu@ustc.edu.cn}
\orcid{0000-0002-3507-9607}
\affiliation{%
  \institution{University of Science and Technology of China \& State Key Laboratory of Cognitive Intelligence}
  \city{Hefei}
  \country{China}
}

\author{Enhong Chen}
\email{cheneh@ustc.edu.cn}
\orcid{0000-0002-4835-4102}
\affiliation{%
  \institution{University of Science and Technology of China \& State Key Laboratory of Cognitive Intelligence}
  \city{Hefei}
  \country{China}
}

\begin{abstract}
\input{content/0.abstract}
\end{abstract}




\maketitle

\section{INTRODUCTION}\label{Sec: intro}
\input{content/1.Introduction}

\section{RELATED WORKS}
\input{content/2.related_work}

\section{PROBLEM DEFINITION}\label{Sec: problem_definition}
\input{content/3.preliminary}

\section{METHODOLOGY}
\input{content/4.method}

\section{EXPERIMENTAL EVALUATION}
\input{content/5.experiments}

\section{CONCLUSIONS}
\input{content/6.conclusion}

\bibliographystyle{ACM-Reference-Format}
\balance
\bibliography{content/references}

\appendix
\section{APPENDIX}
\input{content/7.appendix}

\end{document}

%% file: content/0.abstract.tex
The sequential recommender (SR) system is a crucial component of modern recommender systems, as it aims to capture the evolving preferences of users. Significant efforts have been made to enhance the capabilities of SR systems. These methods typically follow the \textbf{model-centric} paradigm, which involves developing effective models based on fixed datasets. However, this approach often overlooks potential quality issues and flaws inherent in the data. Driven by the potential of \textbf{data-centric} AI, we propose a novel data-centric paradigm for developing an ideal training dataset using a model-agnostic dataset regeneration framework called DR4SR. This framework enables the regeneration of a dataset with exceptional cross-architecture generalizability. Additionally, we introduce the DR4SR+ framework, which incorporates a model-aware dataset personalizer to tailor the regenerated dataset specifically for a target model. To demonstrate the effectiveness of the data-centric paradigm, we integrate our framework with various model-centric methods and observe significant performance improvements across four widely adopted datasets. Furthermore, we conduct in-depth analyses to explore the potential of the data-centric paradigm and provide valuable insights. The anonymous code can be found at \textcolor{blue}{\url{https://github.com/USTC-StarTeam/DR4SR}}.

%% file: content/1.Introduction.tex
\begin{figure}
    \centering
    \includegraphics[scale=0.35]{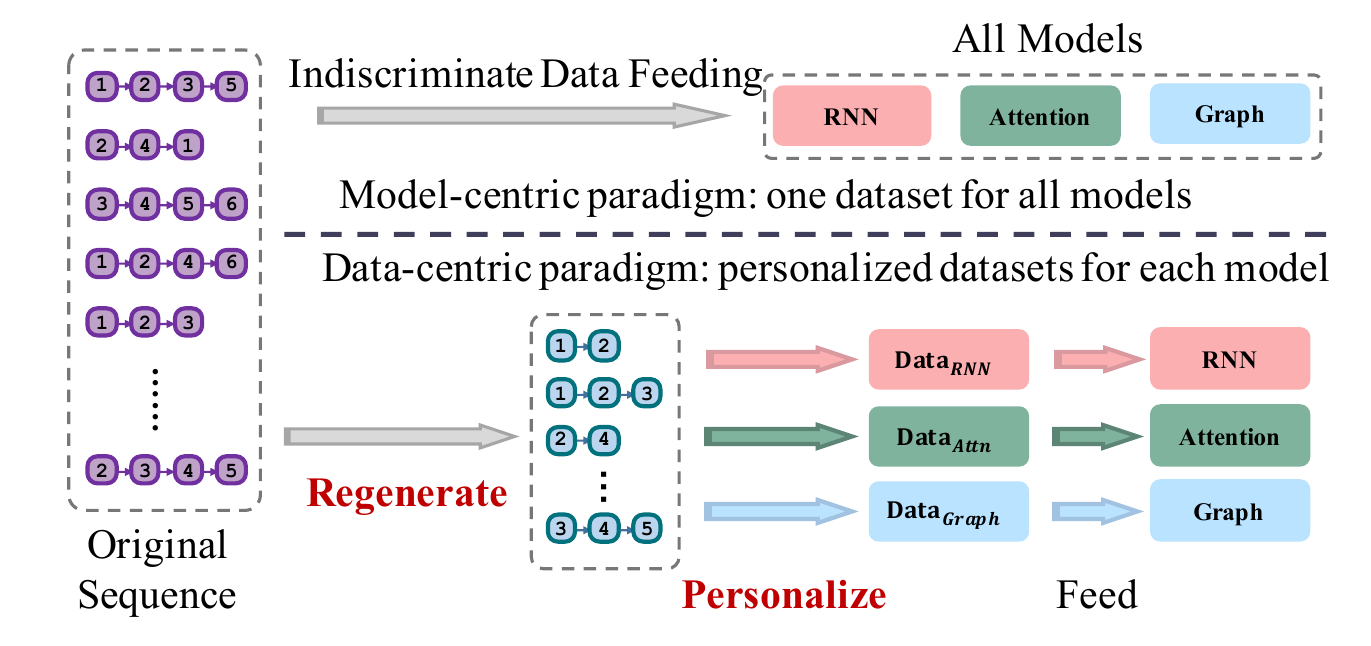}
    \vspace{-0.3cm}
    \caption{Model-centric paradigm v.s. Data-centric pradigm.}
    \label{fig: motivation}
    \vspace{-0.3cm}
\end{figure}


The sequential recommender (SR) system is a critical component of modern recommender systems, aiming to capture the evolving preferences of users through their sequential interaction records \cite{survey_SR1, survey_LLM4Rec}. In recent years, significant efforts have been made to enhance SR. This includes the development of sophisticated deep models \cite{Caser, GRU4Rec, SASRec, SRGNN, UniSRec, RecFormer, ca-cdsr}, effective training strategies \cite{RSS, CT4Rec}, and the refinement of representation space through self-supervised learning \cite{survey_selfrec, S3Rec, CL4SRec, DuoRec, ICLRec, APGL4SR, luankang}, among others. These methods follow the \textbf{model-centric} paradigm, which specifically aims to develop more effective models given fixed datasets.


Despite the remarkable achievements, these methods often overlook potential quality issues inherent within the data \cite{survey_data_centric}, which may lead to overfitting or amplification of data errors \cite{robust_cf, survey_data_centric_RS}. To address these challenges, the \textbf{data-centric} paradigm \cite{survey_data_centric, survey_data_centric_graph, survey_data_centric_RS} has been proposed, focusing on developing high-quality data with fixed models. For example, graph structure learning \cite{survey_GSL, gsl1, gsl2, CCGC} has been introduced to uncover valuable graph structures from data \cite{survey_data_centric_graph}. Additionally, generative models such as GAN \cite{GAN}, VAE \cite{VAE}, and Diffusion model \cite{Diffusion} have been employed to synthesize new training samples \cite{intro_synthesis1, intro_synthesis2}. Inspired by the potential of data-centric AI, we aim to acquire an informative and generalizable training dataset for sequential recommender systems. We define this as the \textbf{training data development} problem. To the best of our knowledge, we are the first to investigate this problem in the context of sequential recommendation systems.


To address the issue, dataset generation is the most relevant topic. However, to the best of our knowledge, it has been rarely studied in the field of recommender systems. UPC-SDG\cite{privacy_data_generation} proposed to generate a synthetic dataset, but its primary goal is to preserve privacy rather than recommendation performance. Another potential solution is dataset distillation (DD), aiming to derive a smaller synthetic dataset and enable trained models to attain comparable performance\cite{survey_DD1, Infinite-AE, DConRec, CGM, TF-DCond}. However, they prioritize training efficiency over effectiveness, which differs from our motivation. Notably, recent advancements in DD \cite{dataset_quant, DATM} have reported that even increasing synthetic dataset size, the performance will ultimately plateau around the optimal performance achieved with the original dataset. Yet another viable approach is denoising sequential recommendation, which involves the removal of noisy information from the original data \cite{CLEA, RAP, DSAN, FMLP, DPT}. However, denoising is just one aspect of the larger issue concerning training data development. Our objective is to excavate innovative data content and formats within the developed data, further enhancing model training.

To acquire the optimal training data, our key idea is to learn a new dataset that explicitly contains item transition patterns. In detail, we decompose the modeling process of a recommender into two stages: extracting transition patterns $\mathcal{X}'$ from the original dataset $\mathcal{X}$ and learning user preferences $\mathcal{Y}$ based on $\mathcal{X}'$. The learning of the mapping $\mathcal{X} \rightarrow \mathcal{Y}$ is challenging since it involves two implicit mappings: $\mathcal{X} \rightarrow \mathcal{X}'$ and $\mathcal{X}' \rightarrow \mathcal{Y}$. Therefore, we explore the possibility of developing a dataset that explicitly represents item transition patterns of $\mathcal{X}'$. This allows us to explicitly decompose the learning process into two phases, with $\mathcal{X}' \rightarrow \mathcal{Y}$ being intuitively easier to learn. Hence, our main focus is on learning an effective mapping function for $\mathcal{X} \rightarrow \mathcal{X}'$, which is a one-to-many mapping. We define the learning process as a \textbf{dataset regeneration} paradigm as depicted in Figure \ref{fig: motivation}, where "re-" indicates that we do not incorporate any additional information but solely rely on the original dataset.

To realize dataset regeneration, we propose a novel data-centric paradigm \textbf{D}ataset \textbf{R}egeneration for \textbf{S}equential \textbf{R}ecommendation (\textbf{DR4SR}), to regenerate the original dataset into a highly informative and generalizable dataset. In detail, we first construct a pre-training task, which makes it possible to perform dataset regeneration. Subsequently, a diversity-promoted regenerator is proposed to model the one-to-many relationship between sequences and patterns in the regeneration process. Lastly, we propose a hybrid inference strategy to regenerate a new dataset with balanced exploration and exploitation. To confirm the superiority of DR4SR, we integrated our framework with various model-centric methods and conducted experiments on four widely adopted datasets. The experimental results can demonstrate the ideal cross-architecture generalizability of DR4SR and the high complementarity of data-centric and model-centric paradigms.

Upon regenerating the dataset, we encounter a novel challenge: the dataset regeneration process operates independently of the target model. While it demonstrates promising cross-architecture generalizability, the regenerated dataset may be sub-optimal for a particular target model. Consequently, our objective is to further customize the regenerated dataset to a particular target model. Nevertheless, the non-differentiable nature of the hybrid inference process poses a difficulty, as optimizing the dataset regenerator based on the downstream recommendation performance via gradient backpropagation becomes unfeasible. To mitigate the aforementioned challenges, we augment DR4SR to a model-aware dataset regeneration process, denoted as DR4SR+. DR4SR+ takes into account the unique attributes of each target model and modifies the regenerated dataset accordingly, as depicted in Figure \ref{fig: motivation}. In particular, we have implemented a dataset personalizer that assigns a score to every pattern within the regenerated dataset. To prevent model collapse, we formulate the optimization of the dataset personalizer as a bi-level optimization problem, which can be efficiently addressed using implicit differentiation. Empirical results substantiated that DR4SR+ can enhance the regenerated dataset further. Additionally, our investigations suggest that the data forms amenable to regeneration are not confined to sequences. Our contributions are summarized as follows:


\begin{itemize}[leftmargin=*]
    \item To the best of our knowledge, this study is the first to address the issue of training data development for sequential recommendations from a data-centric perspective, intending to create an informative and generalizable training dataset.

    \item We propose a model-agnostic dataset regeneration framework DR4SR to tackle the problem. The framework comprises a pre-training task, a diversity-promoted regenerator, and a hybrid inference strategy, all of which contribute to the regeneration of an informative and generalizable dataset.

    \item We further develop DR4SR into a model-aware version, denoted as DR4SR+, to tailor the regenerated dataset to various target models. This is accomplished through the use of a dataset personalizer, which assigns scores to each regenerated pattern and can be optimized efficiently using implicit differentiation.

    \item We integrated DR4SR and DR4SR+ with various model-centric methods, resulting in substantial improvements on four widely adopted datasets. Additionally, we conducted comprehensive analyses to explore the potential of the data-centric paradigm.
\end{itemize}

%% file: content/2.related_work.tex
\subsection{Sequential Recommendation}

To reveal the dynamic preferences embedded within users' interaction sequences, sequential recommendation (SR) has become a prominent branch within the realm of recommendation systems\cite{survey_SR1, survey_SR_lab, ubcs, HyperSoRec, MCNE}. Significant efforts have been directed towards enhancing the capabilities of sequential recommender systems. Initial endeavors concentrated on developing intricate sequential deep models to capture complex sequential preferences, encompassing CNN\cite{Caser, CosRec}, RNN\cite{GRU4Rec, NARM, GUESR}, GNN\cite{SRGNN, GCSAN, GCEGNN}, and Transformer\cite{transformer}-based approaches\cite{SASRec, Bert4Rec}. Beyond model design, ongoing efforts aim to augment the efficacy of SR by devising superior training strategies, such as RSS\cite{RSS} and CT4Rec\cite{CT4Rec}. Additionally, self-supervised learning (SSL) has been incorporated into SR to refine the representation space with diverse SSL tasks, including sequence-level tasks\cite{CL4SRec, ICLRec}, model-level tasks\cite{DuoRec}, and hybrid tasks\cite{S3Rec}.

Recognizing the presence of noise in sequential data, denoising sequential recommendation has emerged as an effective method to filter out noisy information from the original data (e.g., \cite{CLEA, RAP, DSAN, FMLP, DPT, AutoDenoise, end4rec}). One approach achieves this implicitly in the representation space through elaborate model architectures, such as the sparse attention network in DSAN\cite{DSAN}, and the filter-enhanced MLP in FMLP\cite{FMLP}. Another approach explicitly filters out irrelevant items for a target item, including CLEA\cite{CLEA} and RAP\cite{RAP}.


Denoising-based methods partially align with our objective, but denoising alone falls short of producing an optimal dataset as it only focuses on removing information from the data. Dataset generation should produce diverse and novel samples to alleviate the constraints imposed by the original dataset \cite{survey_data_generation}.

\subsection{Data-Centric AI in Recommender System}

Recently, the importance of data in the field of AI has been significantly magnified, culminating in the emergence of the Data-Centric AI (DCAI) paradigm. DCAI represents an all-encompassing concept that comprises three principal components: training data development, inference data development, and data maintenance \cite{survey_data_centric, survey_data_centric_RS}. Within these significant themes, our primary emphasis is on the development of training data, specifically, the creation of an optimal training dataset for sequential recommendations.

In pursuit of the objective of training data development, dataset generation is the most relevant topic. In the field of recommender systems, UPC-SDG\cite{privacy_data_generation} proposed to develop a privacy-preserving synthetic dataset. GTN~\cite{GTN} generated adaptive graph data for graph collaborative filtering. For sequential recommendation, ASReP~\cite{ASReP} and METL~\cite{MELT} focused on generating fabricated data for long-tailed sequences. Another promising technique in DCAI is Dataset Distillation (DD). DD aims to distill a condensed dataset from the original dataset, enabling trained models to attain comparable performance\cite{survey_DD1}. DD is predominantly popular in the field of computer vision (CV) and can be categorized into three main classes: performance matching\cite{performance_matching1}, parameter matching\cite{paramter_matching1}, and distribution matching\cite{distribution_matching1}. Specifically, DD has been introduced to the field of recommender systems by some pioneer works\cite{Infinite-AE, DConRec, CGM, TF-DCond, farzi_data}. $\infty$-AE\cite{Infinite-AE} adopted neural tangent kernel (NTK) to approximate an infinitely wide autoencoder and synthesized fake users with sampling-based reconstruction. DConRec\cite{DConRec} proposed to distill a synthesized dataset by sampling user-item pairs from a learnable probabilistic matrix. Farzi~\cite{farzi_data} achieved efficient sequential data distillation within the latent space. Different from the previous two performance matching methods, CGM\cite{CGM} followed a parameter matching\cite{DC} paradigm to condense categorical recommendation data in the CTR scenario.


While these methods have achieved remarkable success, they diverge from our specific objectives. Our goal is to utilize the generated dataset to enhance performance, whereas these methods primarily prioritize privacy or efficiency concerns.





%% file: content/3.preliminary.tex
In the domain of sequential recommendation, the primary objective is to model user preferences based on their interaction records and provide the next recommended item for them. We can formally define the problem as follows:

\textbf{Definition1}. \textit{(\textbf{Sequential Recommendation}). In a sequential recommendation system, we denote $\mathcal{U}$ as the user set and $\mathcal{V}$ as the item set, and $|\mathcal{U}|$ and $|\mathcal{V}|$ is the respective number of users and items. The interaction sequences of users are ordered chronologically. We define them as $s_u = [v_1, v_2, \dots, v_t, \dots, v_{|s_u|}]$, where $u \in \mathcal{U}$ is some user, $v_t \in \mathcal{V}$ is one item interacted by the user at the time step $t$, $s_u$ represents the sequence, and $|s_u|$ denotes its length, which has a maximum value of $N$. Then the next item prediction task aims to predict the item at the next time step $v_{|s_u| + 1}$ for each user $u \in \mathcal{U}$.}


In deviating from conventional model-centric approaches, which concentrate on enhancing recommendation performance with fixed datasets, we endeavor to construct an optimal training dataset from a data-centric perspective, which is formalized as a training data development problem as follows:

\textbf{Definition2}. \textit{(\textbf{Training Data Development}). Given an original dataset $\mathcal{X}$, traditional methods aim to learn a target model $f'$ satisfying $\mathcal{Y}=f'(\mathcal{X})$, where $\mathcal{Y}$ is the user preferences. Training data development aims to learn a new informative and generalizable dataset $\mathcal{X}'$ satisfying $\mathcal{Y}=f(\mathcal{X}')$, where the target model $f$ can be easier to learn than $f'$ based on $\mathcal{X}'$.}

Then the next crucial issue is how to guarantee an optimal $\mathcal{X}'$. We aim to accomplish this through a model-agnostic dataset regeneration paradigm, which is explained in detail below:

\textbf{Definition3}. \textit{(\textbf{Dataset Regeneration}). Given a sequence $s_u$ in $\mathcal{X}$, dataset regeneration aims to learn a one-to-many dataset regenerator $f'$ satisfying $\{p_{u1}, p_{u2}, \dots, p_{uK}\} = f'(s_u)$, where $\{p_{u1}, p_{u2}, \dots, p_{uK}\}$ are K informative patterns and $f'$ is learned independently from the target model $f$. By collecting all regenerated patterns, we can obtain the new dataset $\mathcal{X}'$.} 


Upon regenerating a new dataset, there is an imperative need to tailor the regenerated dataset specifically for a target recommendation model, as the regeneration process remains independent of the target model. We hence propose a model-aware dataset regeneration process that can evaluate the score of each data sample for the target model. The formal definition is as follows: 

\textbf{Definition4}. \textit{(\textbf{Model-aware Dataset Regeneration}). Given the regenerated dataset $\mathcal{X}'$, a dataset personalizer $g$ with parameter $\phi$ is used to generate a scoring matrix $\mathbf{W} \in \mathbb{R}^{|\mathcal{X}'|}$ to evaluate the score of each data sample. The learning of $g$ will be guided by a target model $f$, thereby achieving dataset personalization. The dataset-score pair $(\mathcal{X}', \mathbf{W})$ will serve as the personalized dataset.}

In the next section, we will illustrate the details of dataset regeneration and model-aware dataset regeneration.

%% file: content/4.method.tex
\subsection{Overall Framework}
\begin{figure*}
    \centering
    \includegraphics[scale=0.45]{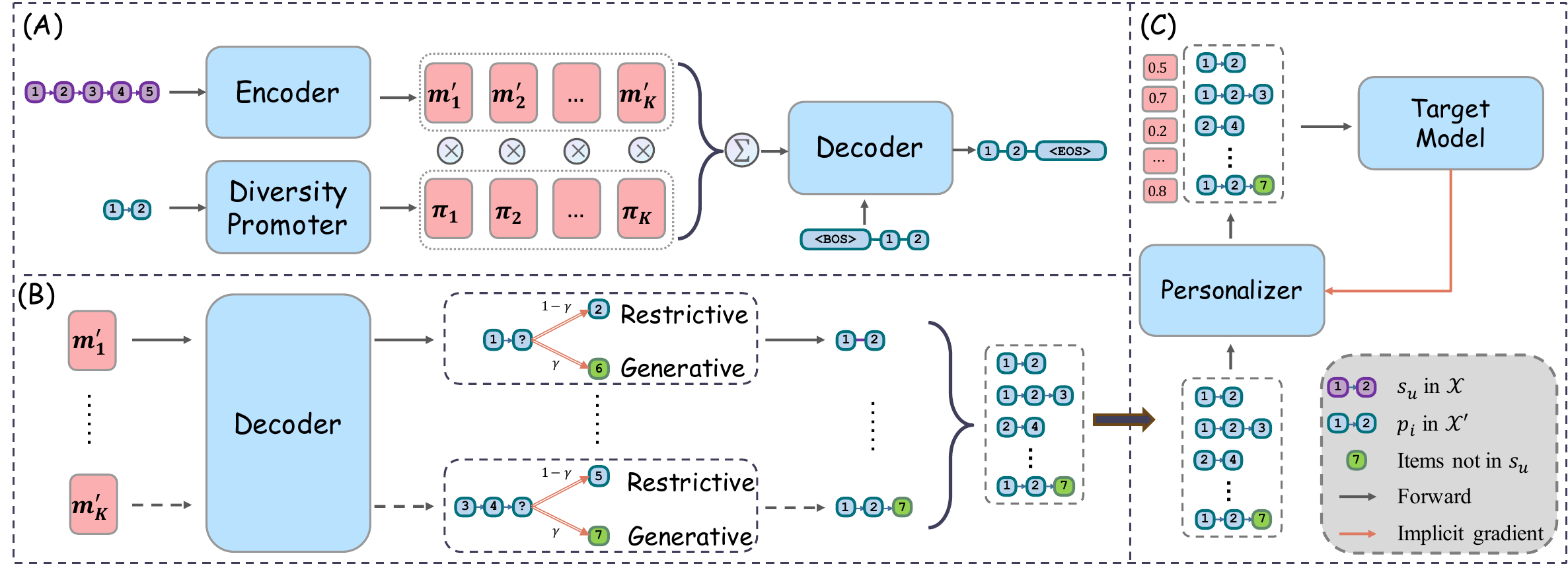}
    \vspace{-0.3cm}
    \caption{The framework of the proposed data-centric paradigm: (A) The Pre-training stage of DR4SR involves training a diversity-promoted data regenerator utilizing the curated pre-training dataset. (B) The inference stage of DR4SR regenerates each source sequence into multiple target patterns with a hybrid inference strategy. (C) Model-aware dataset regeneration with a personalizer further tailors the regenerated dataset to each target model.}
    \label{fig: framework}
    \vspace{-0.2cm}
\end{figure*}

In this paper, we propose a data-centric framework named \textbf{D}ata \textbf{R}egeneration for \textbf{S}equential \textbf{R}ecommendation (\textbf{DR4SR}), aiming to regenerate the original dataset into a highly informative and generalizable dataset, as shown in Figure \ref{fig: framework}. Since the data regeneration process is independent of target models, the regenerated dataset may not necessarily align with their requirements. Therefore, we extend DR4SR to a model-aware version DR4SR+ in Section \ref{Sec: Dataset Personalization}, which tailors the regenerated dataset to a particular target model.


\subsection{Model-agnostic Dataset Regeneration}\label{Sec: Dataset Regeneration}
To develop an informative and generalizable dataset, we aim to construct a dataset regenerator to facilitate the automated dataset regeneration. However, the original dataset lacks supervised information for learning a dataset regenerator. Therefore, we must implement it in a self-supervised learning manner. To achieve this, we introduce a pre-training task in Section \ref{Sec: pre-training task} that can guide the learning of the proposed diversity-promoted regenerator detailed in Section \ref{Sec: diversity-promoted DR}. After completing the pre-training, we proceed to regenerate a new dataset using a hybrid inference strategy in Section \ref{Sec: inference}. The pipeline is detailed in Algorithm \ref{alg: DR4SR}.

\subsubsection{Construction of Pre-training Task}\label{Sec: pre-training task}
\begin{figure}
    \centering
    \includegraphics[scale=0.43]{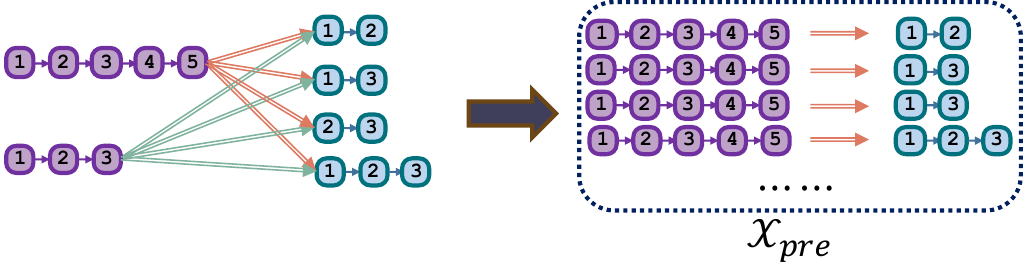}
    \caption{Pre-training task construction. Assuming two given \textcolor{sequences}{sequences} (1,2,3,4,5), (1,2,3) with a window size 3 and a threshold 2, the following \textcolor{patterns}{patterns} can be extracted: (1,2), (1,3), (2,3), (1,2,3). This is because these patterns appear twice within the sliding window. Then the regenerator is supposed to regenerate each sequence into multiple corresponding patterns.}
    \label{fig: pre-training task}
    \vspace{-0.2cm}
\end{figure}

To construct a pre-training task, we first obtain item transition patterns through rule-based methods. Then, the pre-training task is formalized as training the regenerator to learn how to regenerate these patterns from the sequences in the original dataset.

Specifically, we utilize a sliding window strategy to extract patterns. Within a specified sliding window of size $\alpha$, we count the occurrence of item transition patterns. Patterns with an occurrence count exceeding a predetermined threshold $\beta$ will be selected to form the pre-training task. A toy example is depicted in Figure \ref{fig: pre-training task}, where the window size is set to 3 and the threshold is set to 2.

After obtaining the pattern set, one pre-training instance pair can be constructed as $(s_u, p_i)$, where $s_u$ is the interaction sequence of user $u$, $p_i$ is $i$-th pattern extracted from $s_u$. Then the regenerator $f'$ is supposed to regenerate $s_u$ into the corresponding pattern $p_i$ for each pre-training instance pair $(s_u, p_i)$. We denote the entire pre-training dataset as $\mathcal{X}_{\text{pre}}$.

\textbf{Note}: This strategy is related to sequential pattern mining, which aims to uncover patterns for guiding sequential recommendations \cite{survey_RS4, old_pattern_mining}. However, these methods are known to generate redundant information and neglect infrequent patterns, thus failing to meet the requirements of a dataset regenerator.

\subsubsection{Diversity-promoted Regenerator}\label{Sec: diversity-promoted DR}
With the aid of the pre-training task, we can now pre-train a dataset regenerator. In this paper, we employ the Transformer\cite{transformer} model as the main architecture of our regenerator, whose generation capabilities have been extensively validated. The dataset regenerator consists of three modules: an encoder to obtain representations of sequences in the original dataset, a decoder to regenerate the patterns, and a diversity promoter to capture the one-to-many mapping relationship. Next, we will proceed to introduce each of these modules individually. 

The encoder consists of several stacked layers of multi-head self-attention (MHSA) and feed-forward (FFN) layers. Considering a sequence in $\mathcal{X}$, we can get the sequence representation by:
\begin{equation}
    \mathbf{H}^{(l)} = FFN(MHSA(\mathbf{H}^{(l - 1)})),
\end{equation}
where $\mathbf{H}^{(l - 1)} \in \mathbb{R}^{N \times d}$ is the output at the $(l - 1)$-th layer, and $\mathbf{H}^{(0)}$ is item embeddings with added learnable positional encoding. As for the decoder, it takes patterns in the regenerated dataset $\mathcal{X}'$ as input. The objective of the decoder is to reconstruct the pattern given the sequence representations from the encoder:
\begin{equation}\label{eq: reconstruction loss}
    L_{\text{recon}} = - \sum_{(s_u, p_i)}^{|\mathcal{X}_{\text{pre}}|} \sum_{t=1}^{T} \log P(p_{it} | \mathbf{h}^{(l)}_u, \hat{p}_{<t}),
\end{equation}
where $(s_u, p_i)$ is a sequence-pattern pair in $\mathcal{X}_{\text{pre}}$, $t$ is the position in the target pattern, $\mathbf{h}^{(l)}_u$ is the representation of the original sequence $s_u$, $\hat{p}_{<t}$ is the pattern generated before $t$, and $P$ is the prediction probability for the target item $p_{it}$.

However, as mentioned in Section \ref{Sec: pre-training task}, multiple patterns can be extracted from a sequence, which presents challenges during the training process. For instance, a sequence like (1, 2, 3, 4, 5) may yield two distinct patterns, namely (1, 2) and (4, 5). Consequently, this can introduce conflicts during training, thereby impeding the convergence of the dataset regenerator. To address this one-to-many mapping issue, we further propose a diversity promoter.

Specifically, we employ a more aggressive approach to adaptively regulate the impact of the original sequence during the decoding stage by incorporating information about the target pattern. First, we project the memory $\mathbf{m} \in \mathbb{R}^{D}$ (representations of the original sequence) generated by the encoder into K distinct vector spaces, i.e., $\{\mathbf{m}'_1, \mathbf{m}'_2, \dots, \mathbf{m}'_K\}$ and $\mathbf{m}'_k \in \mathbb{R}^{D}$. This projection enables the acquisition of memories with diverse semantic information. Ideally, different target patterns should match distinct memories. To achieve this, we additionally introduce a Transformer encoder to encode the target patterns and obtain $\mathbf{h}^{(l)}_{\text{pattern}} \in \mathbb{R}^{D}$. Notably, it is imperative to exercise caution when incorporating $\mathbf{h}^{(l)}_{\text{pattern}}$ into the decoder, otherwise the model may inadvertently collapse into an identity mapping function by simply duplicating the input. Therefore, we compress $\mathbf{h}^{(l)}_{\text{pattern}}$ into a probability vector by:
\begin{equation}
    \mathbf{\pi} = \text{Softmax}(\text{MLP}(\mathbf{h}^{(l)}_{\text{pattern}})),
\end{equation}
where $\pi = \{\mathbf{\pi}_1, \mathbf{\pi}_2, \dots, \mathbf{\pi}_k, \dots, \mathbf{\pi}_K\}$ and $\mathbf{\pi}_k$ is the probability of choosing the $k$-th memory. To ensure that each memory space receives sufficient training, we do not perform hard selection. Instead, we obtain the final memory through a weighted sum:
\begin{equation}
    \mathbf{m}_{\text{final}} = \sum_{k=1}^{K} \mathbf{\pi}_k \mathbf{m}'_k.
\end{equation}
Ultimately, we can leverage the acquired memory to facilitate the decoding process and effectively capture the intricate one-to-many relationship between sequences and patterns.

\subsubsection{Dataset Regeneration with Hybrid Inference Strategy}\label{Sec: inference}
After pre-training, we utilize the regenerator for inference to obtain a regenerated dataset $\mathcal{X}'$. Specifically, we re-feed the sequences in $\mathcal{X}$ into the regenerator to obtain new patterns. Due to the unavailability of target patterns, we cannot generate a category vector $\pi$ for selecting memories. As a result, we respectively input each memory directly into the decoder, leading to the generation of K patterns. Besides, we consider the following two decoding modes: \textbf{Restrictive mode}. Decoding is limited to selecting items from the input sequence, which focuses on the exploitation of existing information from the current sequence. \textbf{Generative mode}. There are no restrictions, which allows for the exploration of patterns that may not exist in $\mathcal{X}$. For instance, suppose the original dataset contains item transition patterns like (1, 2) and (2, 6). When decoding a sequence like (1, 2, 3, 4, 5), the model may identify a high-order pattern like (1, 2, 6), which was not explicitly present in $\mathcal{X}$.

To achieve a harmonious balance between exploitation and exploration, we propose a hybrid strategy: When decoding an item, there is a probability of $\gamma$ to adopt the generative mode and a probability of $1 - \gamma$ to adopt the restrictive mode.

\subsection{Model-aware Dataset Regeneration}\label{Sec: Dataset Personalization}

Since the previous regeneration process is independent of target models, the regenerated dataset may be sub-optimal for a particular target model. Hence, we extend the model-agnostic dataset regeneration process to a model-aware one. To achieve this, built upon the dataset regenerator, we further introduce a dataset personalizer to evaluate the scores of each data sample in the regenerated dataset in Section \ref{Sec: dataset personalizer}. Then the dataset personalizer can be efficiently optimized by implicit differentiation introduced in Section \ref{Sec: bi-level optimization}. The pipeline is detailed in Algorithm \ref{alg: DR4SR+}.

\subsubsection{Dataset personalizer}\label{Sec: dataset personalizer}

As outlined in Section \ref{Sec: problem_definition}, our objective is to train a dataset personalizer $g$ with parameter $\phi$ that can evaluate the score of each data sample $\mathbf{W}$ for the target model. Specifically, given a data sample of pattern $i$ in the position $t$, the score $w_{i,t}$ for this data sample can be computed as follows:
\begin{equation}
    \begin{gathered}
    \mathbf{z} = g_{\phi}(\mathbf{h}_{t}^i), \\
    \hat{\mathbf{z}} = \text{Softmax}((\mathbf{z} + \mathbf{G})/\tau), \\
    w_{i,t} = \hat{\mathbf{z}}_1,
    \end{gathered}
\end{equation}
where $g_{\phi}$ is the dataset personalizer implemented as an MLP,  $\mathbf{h}_{t}^i$ is the representation of pattern $i$ at position $t$ generated by the target model $f'$, $\mathbf{z} \in \mathbb{R}^2$ is the computed logits vector, Gumbel noise $\mathbf{G}$ is sampled from a Gumbel distribution\cite{gumbel_softmax}, $\tau$ is a temperature parameter to control the smoothness of Gumbel distribution, and the subscript $1$ means selecting the first element of $\hat{\mathbf{z}}$ as the score. Since the personalizer takes pattern representations as input, the personalizer possesses a receptive field over the entire pattern, enabling a comprehensive evaluation of the pattern. Additionally, the Gumbel noise is introduced to encourage the scores to approximate a discrete selection operation, which is a desirable property\cite{gumbel_softmax}.

To ensure the generality of the framework, we utilize the calculated score to adjust the weighting of training losses, which does not necessitate any additional modifications to target models. We start by defining the original next-item prediction loss:
\begin{equation}\label{eq: overall}
    \mathcal{L}_{rec-ori} = \sum_{i = 1}^{|\mathcal{X}'|} \sum_{t=2}^{|p_i|} \mathcal{L}_{next-item}(i, t),
\end{equation}
\begin{equation}\label{eq: rec loss}
    \mathcal{L}_{next-item}(i, t) = -\log(\sigma(\mathbf{h}_{t-1}^i \cdot \mathbf{v}_t^i)) - \sum_{v_j \notin p_i} \log(1 - \sigma(\mathbf{h}_{t-1}^i \cdot \mathbf{v}_j)),
\end{equation}
where $p_i$ represents one regenerated pattern, $\mathbf{h}_{t-1}^i$ is the representation of pattern $i$ at position $t - 1$, $\mathbf{v}_t^i$ is the embedding of the target item in $p_i$, $\sigma$ is the sigmoid function, and we randomly sample one negative item $v_j$ at each time step $t$ for each pattern. The purpose of Equation \ref{eq: rec loss} is to maximize the probability of recommending the expected next item $v_t^i$ given the pattern representation $\mathbf{h}_{t-1}^i$.

Subsequently, the training loss function on the personalized dataset can be defined as follows:
\begin{equation}\label{eq: rec loss with weight}
    \mathcal{L}_{rec} = \sum_{i = 1}^{|\mathcal{X}'|} \sum_{t=2}^{|p_i|} \mathbf{w}_{i,t} \mathcal{L}_{next-item}(i, t).
\end{equation}


\subsubsection{Efficient optimization of dataset personalizer}\label{Sec: bi-level optimization}

The optimization of the dataset personalizer presents a nontrivial challenge. Should we attempt to optimize it concurrently with the target model using Equation \ref{eq: rec loss with weight}, we are likely to confront a model collapse issue, wherein the personalizer assigns zero to each data sample. To confront this challenge, we formalize the optimization process as a bi-level optimization problem.

Specifically, our goal is to train an optimal dataset personalizer to obtain an optimal personalized dataset, on which the target model can achieve the best recommendation performance:
\begin{equation}\label{eq: bi-level optimization}
\begin{gathered}
    \phi^* = \arg \min_{\phi} L_{\text{dev}} (\theta^*(\phi)), \\
    s.t. \quad \theta^*(\phi) = \arg \min_{\theta} L_\text{train}(\theta, \phi),
\end{gathered}
\end{equation}
where $L_\text{train}(\theta, \phi)$ is defined in Equation \ref{eq: rec loss with weight}. The lower optimization aims to get an optimal target model with parameter $\theta^*$ while fixing the dataset personalizer. Besides, $\theta^*$ is an implicit function of $\phi$, so it can be denoted as $\theta^*(\phi)$. $L_{\text{dev}} (\theta^*(\phi))$ is defined in Equation \ref{eq: overall}, which is the unweighted recommendation loss of the target model on a validation dataset. In the context of batch training, we can shuffle the original dataset to create a validation set\cite{BIAO}.

Then we can elaborate on how to optimize the bi-level optimization problem. Theoretically, achieving the optimal point of the target model through training is a prerequisite before conducting a one-step upper-level optimization. For the sake of efficiency, we conduct a one-step upper optimization after only $T_{\text{lower}}$ rounds of lower-level optimization in our implementation. The lower optimization can be directly optimized with gradient descent. As for the upper optimization, we need to compute $\nabla_{\phi} L_{\text{dev}} (\theta^*(\phi))$. Given that $\theta^*(\phi)$ is an implicit function of $\phi$, we can derive:
\begin{equation}\label{eq: final implicit optimization}
    \nabla_{\phi} L_{\text{dev}} = - \nabla_{\theta} L_{\text{dev}} \cdot \sum_{n=0}^K (I - \nabla_{\theta}^2 L_{\text{train}})^n \cdot \nabla_{\phi} \nabla_{\theta} L_{\text{train}}.
\end{equation}
The detailed derivations can be found in Appendix \ref{appendix: derivation}. Equation \ref{eq: final implicit optimization} can be efficiently calculated with vector-Jacobi product\cite{aux_learn2}. Through this, we can personalize the regenerated dataset without incurring a substantial time overhead.


%% file: content/5.experiments.tex

\subsection{Experimental Settings}\label{Sec: experimental setting}
\subsubsection{Datasets}
\input{table/statistics}
To validate the effectiveness of our proposed approach, we follow previous works\cite{CL4SRec, ICLRec} to conduct experiments on four commonly used and publicly available datasets:
\begin{itemize}[leftmargin=*]
\item \textbf{Beauty, Sports, Toys\footnote{\url{http://snap.stanford.edu/data/amazon/productGraph/categoryFiles/}}}: The Amazon-review dataset has emerged as a prominent resource for evaluating recommendation systems. For our study, we specifically choose three categories, namely "Beauty," "Sports and Outdoors," and "Toys and Games."
\item \textbf{Yelp\footnote{\url{https://www.yelp.com/dataset}}}: The Yelp dataset is a widely used business review dataset that contains information about businesses, users, and reviews. 
\end{itemize}

We follow the preprocessing in \cite{CL4SRec, ICLRec} to guarantee a minimum of 5 interactions associated with each user and item.
Statistics of the preprocessed datasets are summarized in Table \ref{tab: dataset}.

\subsubsection{Compared Baselines and Target Models}
To verify the superiority of DR4SR, we select two representative data-centric baselines:
\begin{itemize}[leftmargin=*]
    \item $\infty$-AE~\cite{Infinite-AE}: an AutoEncoder-based framework, synthesizing user interaction data with sampling-based reconstruction.
    \item MELT~\cite{MELT}: it focused on the long-tail problem through data generation for long-tail users and items, respectively.
\end{itemize}
To fully demonstrate the cross-architecture generalizability of DR4SR, we select target models from various categories:
\begin{itemize}[leftmargin=*]
    \item RNN-based: GRU4Rec\cite{GRU4Rec} is a sequential recommender system that applies GRU layers to capture dynamic user preferences.
    \item Transformer-based: SASRec\cite{SASRec} is a sequential recommendation model that leverages self-attention mechanisms\cite{transformer} to make item recommendations for the next user action.
    \item Denoising-based: FMLP\cite{FMLP} is one of the most representative denoising-based SR methods, which adopts a filter-enhanced MLP to remove noises in interaction sequences.
    \item GNN-based: GCE-GNN\cite{GCEGNN} constructs an item graph to capture global context information facilitating recommendation.
    \item Contrastive-learning-based: CL4SRec\cite{CL4SRec} is one of the most powerful contrastive sequential recommenders, which introduces three sequence-level augmentation strategies.
\end{itemize}
The proposed framework is a data-centric framework, which is complementary to those model-centric methods. Therefore, we integrate DR4SR with all of them to validate the cross-architecture generalizability of DR4SR:
\begin{itemize}[leftmargin=*]
    \item Backbone: We train one target model on the original dataset.
    \item DR4SR: We integrate DR4SR with the target model.
    \item DR4SR+: We integrate DR4SR+ with the target model.
\end{itemize}


\subsubsection{Evaluation Protocols}
To assess the performance of the next-item recommendations, we adopt the widely-used leave-one-out strategy\cite{SASRec, CL4SRec, ICLRec}.
Considering evaluation metrics, we employ two ranking metrics Recall@\{10, 20\} and NDCG@\{10, 20\}. Furthermore, we follow the suggestion of Krichene and Rendle\cite{sampled_metrics} to adopt the whole item set as the candidate item set during evaluation.

\subsubsection{Implementation Details}
To ensure a fair comparison environment, we implement DR4SR and all target models based on an open-source recommender system library RecStudio\cite{recstudio}. The training process is configured with a maximum of 1000 epochs, and early stopping is employed if the NDCG@20 on the validation data does not improve for 20 consecutive epochs. The training batch size $B$ is fixed to 256, Adam\cite{Adam} with a learning rate of 1e-3 is used as the optimizer, the embedding size $d$ is set to 64, and the maximum sequence length $N$ is set to 50 for all datasets. BCE loss is used for all baselines. Detailed hyper-parameter settings of each model can be found in Appendix \ref{appendix: hyper-parameter}. Notably, we fixed the hyper-parameters of each target model when integrating DR4SR with them.

\input{table/regenerated_statistics}
\subsection{Overall Performance}\label{Sec: overall performance}
\input{table/overall}

We compared the performance of each target model with "DR4SR" and "DR4SR+" variants to verify the efficacy of the proposed framework. The statistics of regenerated datasets can be found in Table \ref{tab: regenerated_dataset} in the Appendix. From the overall performance presented in Table \ref{tab: overall performance}, we can draw the following conclusions: (1) DR4SR can regenerate an informative and generalizable dataset: Across all datasets and target models, we can observe a considerable performance enhancement when comparing the "DR4SR" variant with target models. (2) Different target models prefer different datasets: An additional improvement can be observed when comparing the "DR4SR" variant with the "DR4SR+" variant. This is because DR4SR is independent of target models, which makes the regenerated dataset sub-optimal. (3) Denoising is just one aspect of the broader training data development problem: We can find that FMLP can yield significant performance improvement when integrated with DR4SR and DR4SR+. This is because dataset regeneration can additionally capture the one-to-many relationship of sequences and patterns. Moreover, dataset regeneration has the potential to discover latent transition patterns with the hybrid inference mode. (4) Data-centric and model-centric paradigms are complementary: The proposed data-centric paradigm has consistently demonstrated benefits across various datasets and target models, emphasizing the complementarity of the data-centric and model-centric paradigms. This underscores the potential for advancing recommender systems from a data-centric perspective.



\subsection{Ablation Study}
\input{table/ablation}


In this section, we aim to evaluate the efficacy of each module within DR4SR. An ablation study is conducted using SASRec as the target model. We have designed the following model variants: (A) "-diversity": This aims to confirm the significance of capturing the one-to-many relationship by replacing the diversity-promoted regenerator with a vanilla transformer. (B) "pattern": This seeks to substantiate the need for a trainable regenerator by employing patterns extracted based on predefined rules as the regenerated dataset. (C) "end-to-end": This endeavors to illustrate the importance of bi-level optimization by optimizing the dataset personalizer in an end-to-end fashion. Results are presented in Table \ref{tab: ablation study}.


From the table, we can obtain the following conclusions: (1) Comparing (A) with "DR4SR+", we can observe significant performance loss, which demonstrates the necessity of modeling the one-to-many mapping relationship between the original sequences and patterns. (2) Comparing (B) and "SASRec", we can observe that using only the pattern dataset leads to a decrease in performance since many valuable infrequent patterns are ignored. Thus highlighting the superiority of the proposed powerful regenerator. (3) Comparing (C) and "DR4SR+", we find that the end-to-end learning approach yields poor performance, which is unsurprising as the end-to-end training may readily converge to a trivial solution where the dataset personalizer assigns zero scores to all data samples.


\subsection{Advanced Study}\label{Sec: advanced study}
To demonstrate the superiority of the proposed paradigm, we have conducted in-depth analysis experiments from various aspects. The obtained experimental results and analysis are as follows:

\subsubsection{\textbf{Analysis of efficiency of dataset personalization}} 

\input{table/runtime}

Bi-level optimization is widely recognized for its inefficiency and high memory consumption\cite{aux_learn2}, which hinders its practical applicability. Consequently, we conducted an efficiency analysis to validate the efficiency of the adopted bi-level optimization process. The experimental results are presented in Table \ref{tab: runtime}. It can be observed from the results that the training of the dataset personalizer using the implicit differentiation method does not incur significant additional time and space overhead. This demonstrates that the regenerated dataset can be efficiently tailored to different target models.

\subsubsection{\textbf{Analysis of data sample scores of different target models}} 
\begin{figure*}
    \centering
    \includegraphics[scale=0.55]{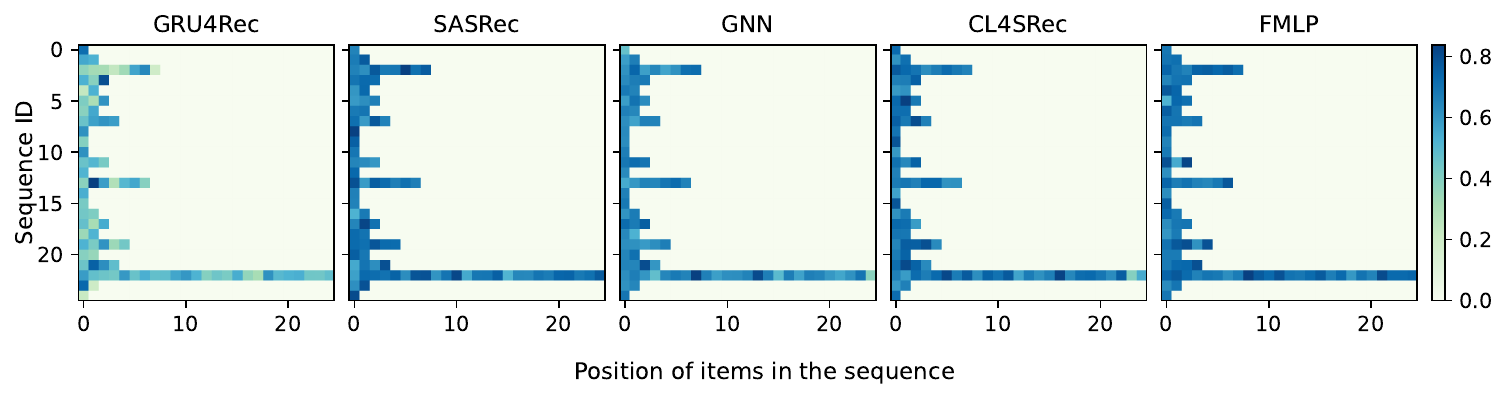}
    \vspace{-0.4cm}
    \caption{Scores assigned by the dataset personalizer for different target models on Toys. From left to right, the variances of the scores for each target model are 0.0280, 0.0244, 0.0241, 0.0248, and 0.0247.}
    \label{fig: loss weight}
    \vspace{-0.2cm}
\end{figure*}

We randomly selected 25 sequences from the regenerated dataset and visualized the sample scores learned by different target models, as depicted in Figure \ref{fig: loss weight}. The figure reveals noticeable variations in dataset preferences among different models. Notably, GRU4Rec demonstrates a higher degree of score variance compared to other methods, which exhibit more uniform weight assignments. This is because the powerful self-attention mechanism or filter-enhanced MLP can adaptively determine the utilization of items in the sequence to some extent. However, even among these methods, discernible preference differences can still be observed. These observations further validate the crucial role of the dataset personalizer.

\begin{figure}
    \centering
    \begin{subfigure}[t]{0.22\textwidth}
           \centering
           \includegraphics[width=\textwidth]{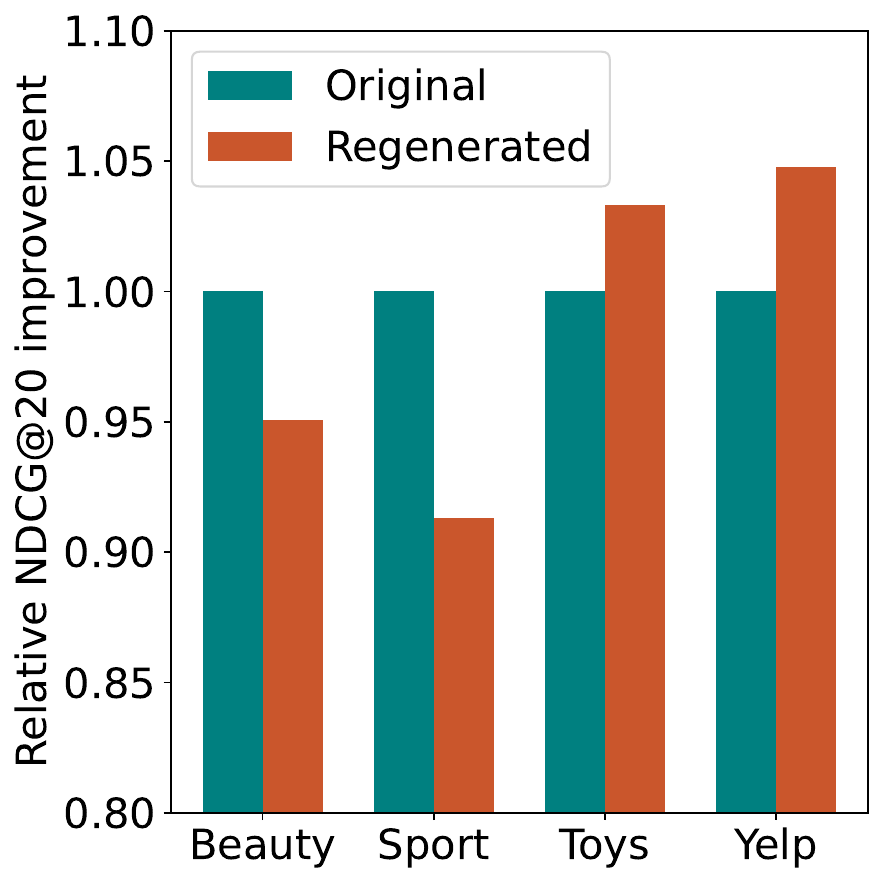}
            \caption{Different graphs}
            \label{fig: data_form_analysis_graph}
    \end{subfigure}
    \begin{subfigure}[t]{0.22\textwidth}
            \centering
            \includegraphics[width=\textwidth]{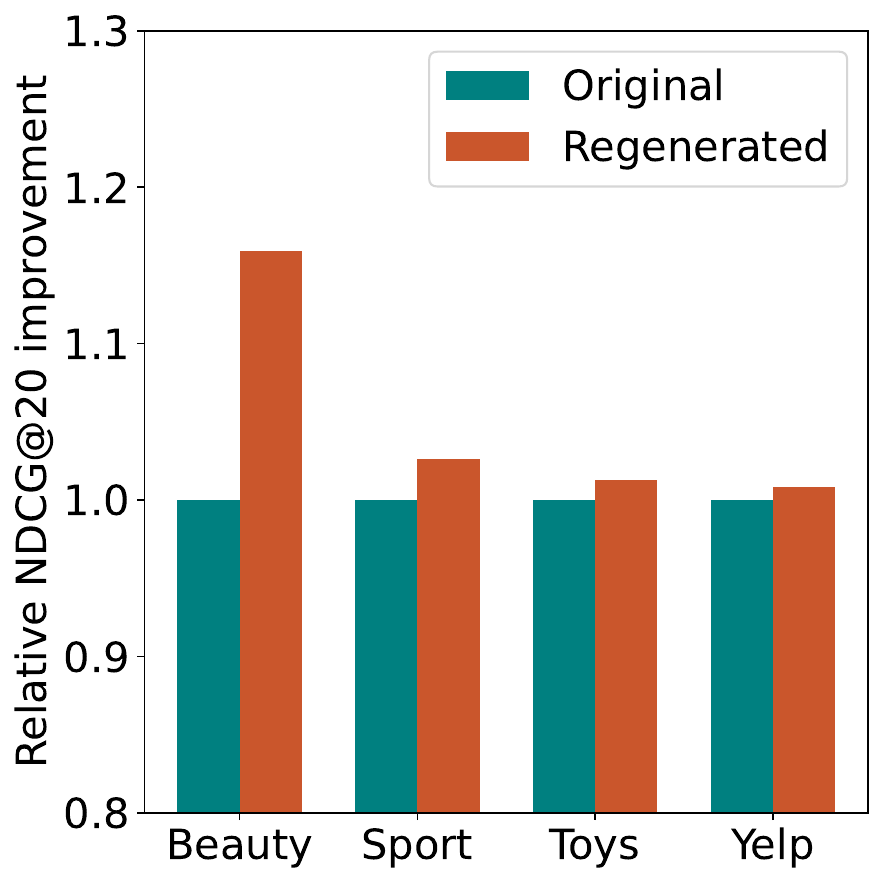}
            \caption{Different augmentations}
            \label{fig: data_form_analysis_aug}
    \end{subfigure}
    \vspace{-0.2cm}
    \caption{Relative NDCG@20 improvement of graphs and data augmentations on different datasets.}
    \label{fig: K analysis}
    \vspace{-0.3cm}
\end{figure}

\subsubsection{\textbf{Analysis of data forms that should be regenerated.}}\label{Sec: beyond sequence}
After regenerating a dataset, we can construct a new graph for GNN and perform data augmentation based on the new dataset. This naturally raises the question of which dataset should be adopted for graph construction or data augmentation: the original or the regenerated one. To shed light on this matter, we further conducted two experiments. Firstly, we compared the effects of using different graphs, and the results are shown in Figure \ref{fig: data_form_analysis_graph}. The results are subtle, with the original graph performing better on half of the dataset, while the regenerated graph yields superior results on the other half. Subsequently, we compared the effects of using different augmentation data, and the results are shown in Figure \ref{fig: data_form_analysis_aug}. From the table, we can observe that DR4SR yields superior data augmentation samples for CL4SRec on the first two datasets. This is attributed to the regenerated sequences containing richer semantic information. However, for the last two datasets, we notice that there is no significant improvement.

Considering these results, we can observe that while DR4SR can provide curated sequence training datasets for diverse target models, certain advanced data formats such as graphs and data used for augmentation still need to be constructed based on the original dataset in certain scenarios. Therefore, we conclude that for the sequential recommendation, the scope of data regeneration can be extended beyond the interaction sequences. For instance, graph structure learning\cite{survey_GSL} can be employed to construct a more suitable correlation graph, and leverage learnable data augmentation\cite{learnable_augmentation1, learnable_augmentation2, CONVERT, DealMVC} can be utilized for contrastive learning. In this study, we primarily focus on regenerating sequences, and the exploration of regenerating other data forms will be addressed in future work.

\subsection{Hyper-parameter Sensitivity}\label{Sec: parameter sensitivity}
In this subsection, we investigate the hyper-parameter sensitivity of the proposed paradigm. Specifically, we focus on the impact of two parameters on model performance, the diversity factor $K$ and the generative decoding probability $\gamma$. For the sake of simplicity, we conducted experiments on Toys and Beauty based on SASRec.

\subsubsection{The diversity factor K}

\begin{figure}
    \centering
    \begin{subfigure}[t]{0.21\textwidth}
           \centering
           \includegraphics[width=\textwidth]{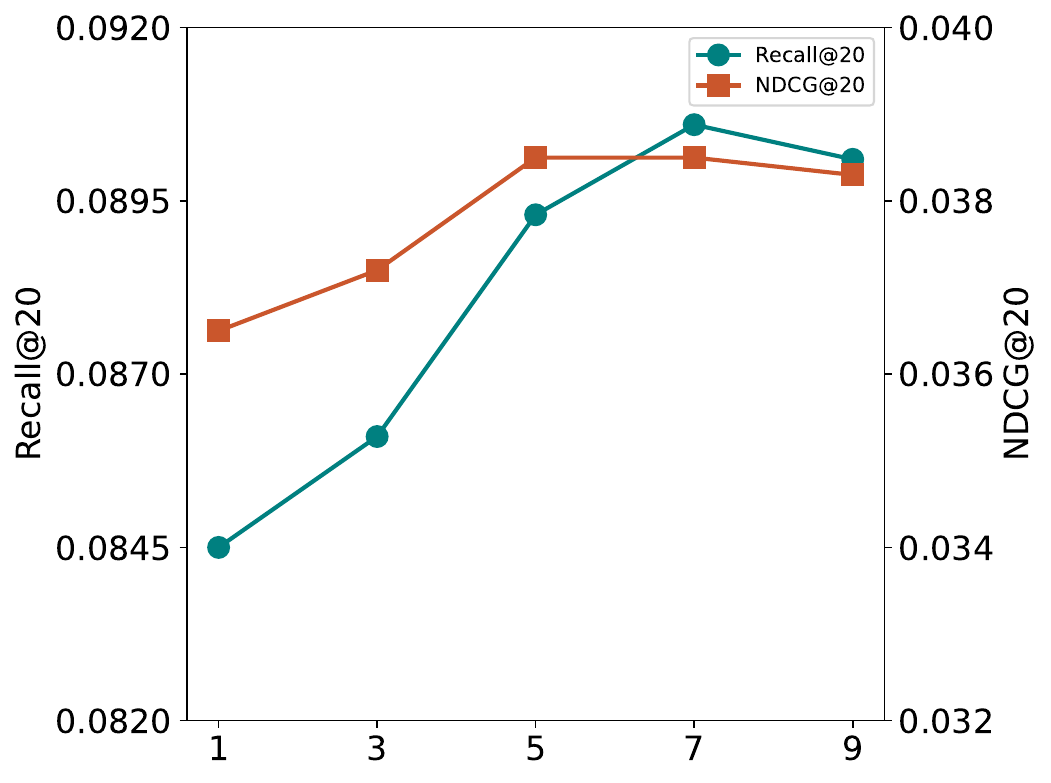}
            \caption{Beauty}
            \label{fig: K_analyasis_beauty}
    \end{subfigure}
    \begin{subfigure}[t]{0.21\textwidth}
            \centering
            \includegraphics[width=\textwidth]{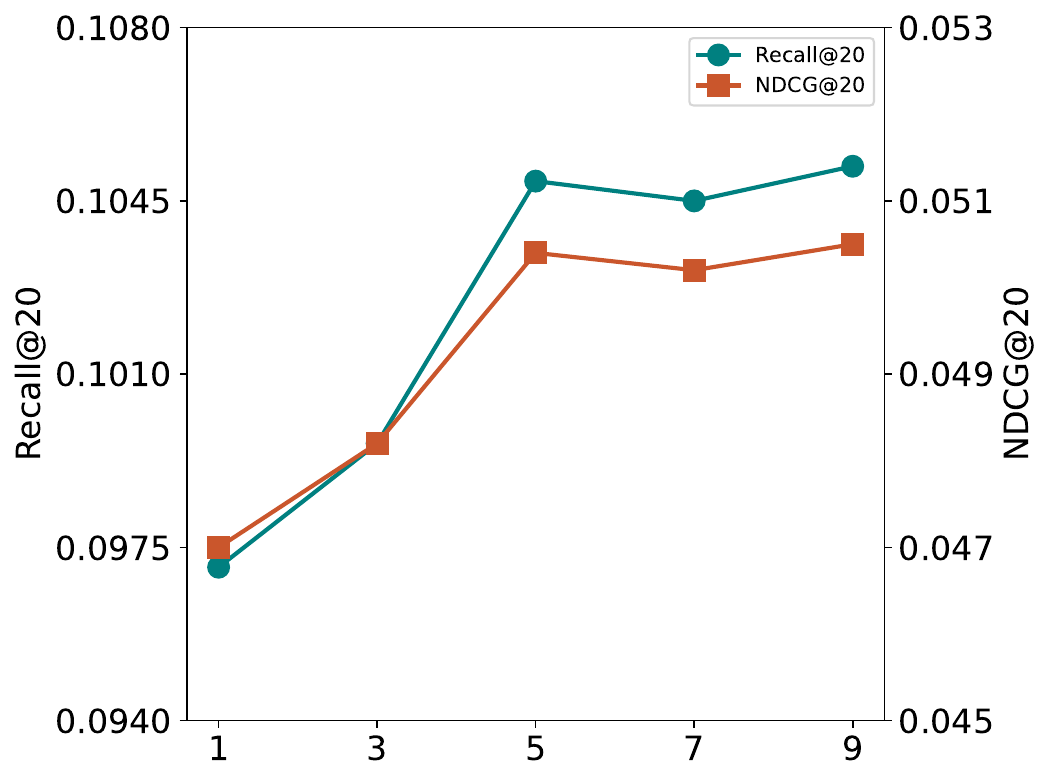}
            \caption{Toys}
            \label{fig: K_analyasis_toy}
    \end{subfigure}
    \vspace{-0.3cm}

    \caption{Recommendation Performance w.r.t different K.}
    \label{fig: K analysis}
    \vspace{-0.2cm}
\end{figure}

We set K among $[1, 3, 5, 7, 9]$ and the results are depicted in Figure \ref{fig: K analysis}, where the performance exhibits an increasing trend as K increases. This is because by increasing K, we can effectively reduce confusion during training and simultaneously enhance the diversity of the regenerated data. Furthermore, once K exceeded a certain threshold, further increasing did not lead to significant changes. This indicates that the regenerator had already captured all available patterns, reaching a plateau phase.

\subsubsection{The generative decoding probability $\gamma$}

\begin{figure}
    \centering
    \begin{subfigure}[t]{0.21\textwidth}
           \centering
           \includegraphics[width=\textwidth]{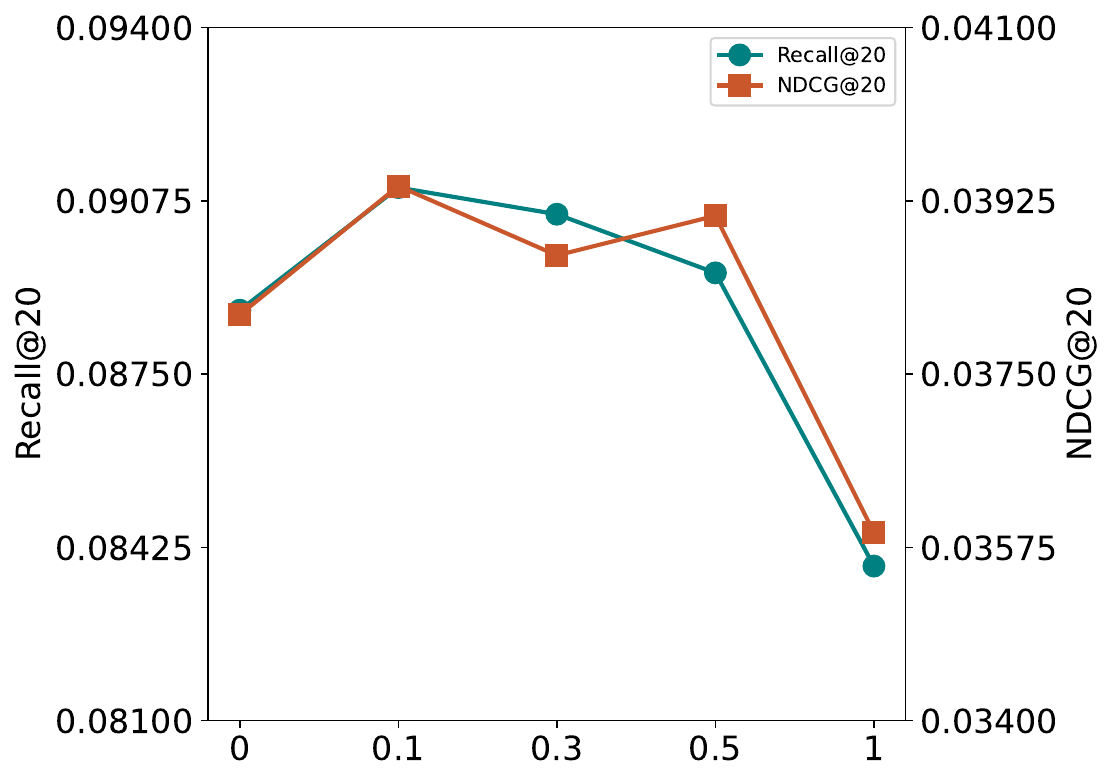}
            \caption{Beauty}
            \label{fig: gamma_analysis_beauty}
    \end{subfigure}
    \begin{subfigure}[t]{0.21\textwidth}
            \centering
            \includegraphics[width=\textwidth]{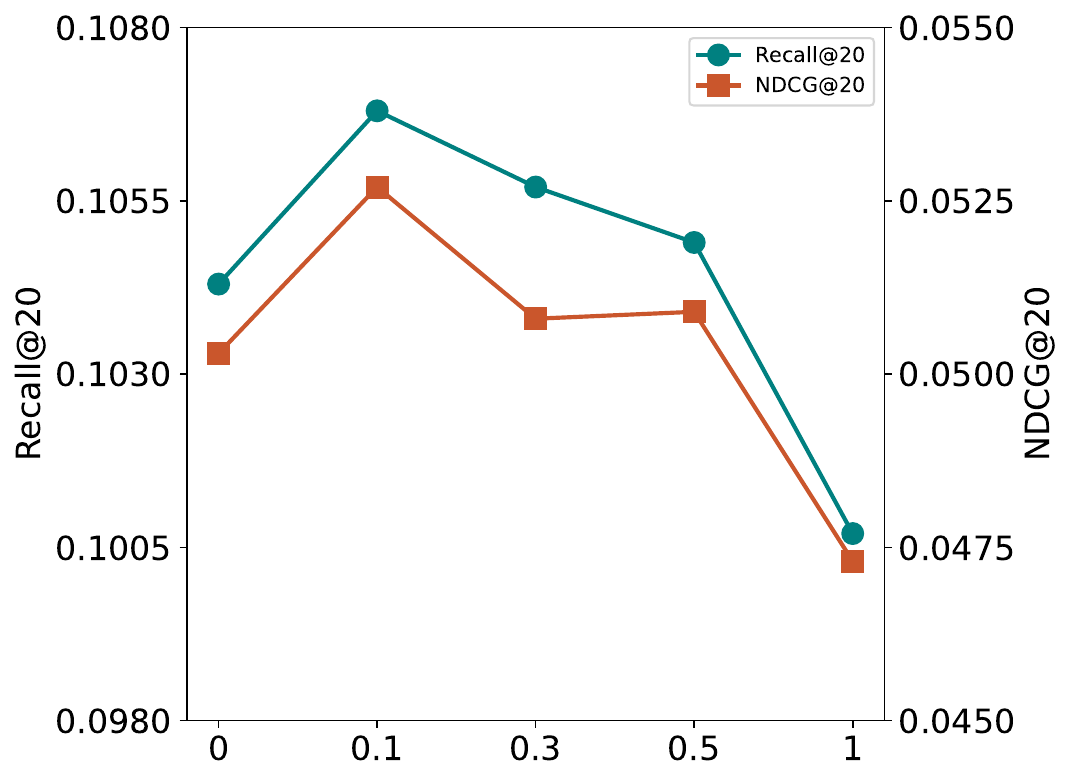}
            \caption{Toys}
            \label{fig: gamma_analysis_toy}
    \end{subfigure}
    \vspace{-0.3cm}

    \caption{Recommendation Performance w.r.t different $\gamma$.}
    \label{fig: gamma analysis}
    \vspace{-0.2cm}
\end{figure}

We set the parameter $\gamma$ within the range of $[0, 0.1, 0.3, 0.5, 1]$ and visualized the results in Figure \ref{fig: gamma analysis}. The performance initially shows an increasing trend followed by a decrease. This observation highlights the importance of striking a balance between exploration and exploitation during the regeneration process. When $\gamma$ is too low, the regenerator is unable to sufficiently explore higher-order information. Conversely, when $\gamma$ is set too high, the model underutilizes the existing information, leading to a decrease in the reliability of the generated results.

%% file: table/statistics.tex
\begin{table}
\centering
\caption{Statistics of the datasets.}
\label{tab: dataset}
\vspace{-0.2cm}
\begin{tabular}{l|llll} 
\hline
Dataset         & Beauty  & Sports  & Toys    & Yelp     \\ 
\hline
$|\mathcal{U}|$ & 22,363  & 35,598  & 19,412  & 30,431   \\
$|\mathcal{V}|$ & 12,101  & 18,357  & 11,924  & 20,033   \\
\# Interactions & 0.2m    & 0.3m    & 0.17m   & 0.3m     \\
Avg. length     & 8.9     & 8.3     & 8.6     & 8.3      \\
Sparsity        & 99.95\% & 99.95\% & 99.93\% & 99.95\%  \\
\hline
\end{tabular}
\end{table}

%% file: table/regenerated_statistics.tex
\begin{table}
\centering
\caption{Statistics of the regenerated datasets.}
\label{tab: regenerated_dataset}
\vspace{-0.2cm}
\begin{tabular}{l|llll}
\hline
Dataset        & Beauty  & Sport   & Toys    & Yelp    \\ \hline
\#users        & 22,363  & 35598   & 19412   & 30431   \\
\#items        & 12101   & 18357   & 11924   & 20033   \\
\#interactions & 0.32m   & 0.45m   & 0.30m   & 0.52m   \\
\#Avg.length   & 4.0     & 4.2     & 3.7     & 4.1     \\
Sparsity       & 99.87\% & 99.93\% & 99.87\% & 99.91\% \\ \hline
\end{tabular}
\end{table}

%% file: table/overall.tex
\begin{table*}[]
\caption{The overall performance. Considering a target model, the best result is bolded while the second-best result is underlined. Superscript * means improvements are statistically significant with p<0.05 while ** meaning p<0.01.}
\label{tab: overall performance}
\vspace{-0.2cm}
\resizebox{\textwidth}{!}{
\begin{tabular}{l|cccc|cccc|cccc|cccc}
\hline
Dataset & \multicolumn{4}{c|}{Beauty}                                           & \multicolumn{4}{c|}{Sports}                                           & \multicolumn{4}{c|}{Toys}                                             & \multicolumn{4}{c}{Yelp}                                              \\ \hline
Metric  & R@10            & R@20            & N@10            & N@20            & R@10            & R@20            & N@10            & N@20            & R@10            & R@20            & N@10            & N@20            & R@10            & R@20            & N@10            & N@20            \\ \hline
$\infty$-AE & 0.0478          & 0.0661          & 0.0262          & 0.0308          & 0.0256          & 0.0373          & 0.0144          & 0.0173          & 0.0450          & 0.0593          & 0.0268          & 0.0304          & 0.0252          & 0.0424          & 0.0121          & 0.0162          \\
MELT        & 0.0577          & 0.0879          & 0.0303          & 0.0379          & 0.0311          & 0.0488          & 0.0163          & 0.0208          & 0.0709          & 0.0987          & 0.0401          & 0.0473          & 0.0293          & 0.0497          & 0.0143          & 0.0195          \\ \hline \hline

GRU4Rec & 0.0204          & 0.0382          & 0.0107          & 0.0150          & 0.0160          & 0.0279          & 0.0085          & 0.0115          & 0.0212          & 0.0357          & 0.0099          & 0.0136          & 0.0215          & 0.0364          & 0.0105          & 0.0143          \\
DR4SR   & \underline{0.0252$^{**}$} & \underline{0.0448$^{**}$} & \underline{0.0128$^{**}$} & \underline{0.0177$^{**}$} & \underline{0.0208$^{**}$} & \underline{0.0341$^{**}$} & \underline{0.0102$^{**}$} & \underline{0.0135$^{**}$} & \underline{0.0252$^{**}$} & \underline{0.0418$^{**}$} & \underline{0.0124$^{**}$} & \underline{0.0165$^{**}$} & \underline{0.0235$^{**}$} & \underline{0.0403$^{**}$} & \underline{0.0114$^{**}$} & \underline{0.0156$^{**}$} \\
Improv  & 23.5\%          & 17.3\%          & 19.6\%          & 18.0\%          & 30.0\%          & 22.2\%          & 20.0\%          & 17.4\%          & 18.9\%          & 22.4\%          & 25.3\%          & 21.3\%          & 9.30\%          & 10.7\%          & 8.57\%          & 9.09\%          \\
DR4SR+  & \textbf{0.0292$^{**}$} & \textbf{0.0473$^{**}$} & \textbf{0.0149$^{**}$} & \textbf{0.0194$^{**}$} & \textbf{0.0223$^{**}$} & \textbf{0.0360$^{**}$} & \textbf{0.0116$^{**}$} & \textbf{0.0151$^{**}$} & \textbf{0.0274$^{**}$} & \textbf{0.0456$^{**}$} & \textbf{0.0134$^{**}$} & \textbf{0.0179$^{**}$} & \textbf{0.0243$^{**}$} & \textbf{0.0415$^{**}$} & \textbf{0.0120$^{**}$} & \textbf{0.0164$^{**}$} \\
Improv  & 43.1\%          & 23.8\%          & 39.3\%          & 29.3\%          & 39.4\%          & 29.0\%          & 36.5\%          & 31.3\%          & 29.2\%          & 27.7\%          & 35.4\%          & 31.6\%          & 13.0\%          & 14.0\%          & 14.3\%          & 14.7\%          \\ \hline
SASRec  & 0.0553          & 0.0847          & 0.0291          & 0.0368          & 0.0297          & 0.0449          & 0.0156          & 0.0194          & 0.0682          & 0.0951          & 0.0381          & 0.0448          & 0.0289          & 0.0488          & 0.0143          & 0.0193          \\
DR4SR   & \underline{0.0595$^{**}$} & \underline{0.0906$^{**}$} & \underline{0.0317$^{**}$} & \underline{0.0395$^{**}$} & \underline{0.0330$^{**}$} & \underline{0.0512$^{**}$} & \underline{0.0174$^{**}$} & \underline{0.0220$^{**}$} & \underline{0.0762$^{**}$} & \underline{0.1049$^{**}$} & \underline{0.0432$^{**}$} & \underline{0.0504$^{**}$} & \underline{0.0304$^{*}$} & \underline{0.0512$^{*}$} & \underline{0.0151$^{*}$} & \underline{0.0202$^{*}$} \\
Improv  & 7.59\%          & 6.97\%          & 8.93\%          & 7.34\%          & 11.1\%          & 14.0\%          & 11.5\%          & 13.4\%          & 11.7\%          & 10.3\%          & 13.4\%          & 12.5\%          & 5.19\%          & 4.92\%          & 5.59\%          & 4.66\%          \\
DR4SR+  & \textbf{0.0619$^{**}$} & \textbf{0.0919$^{**}$} & \textbf{0.0337$^{**}$} & \textbf{0.0412$^{**}$} & \textbf{0.0349$^{**}$} & \textbf{0.0525$^{**}$} & \textbf{0.0191$^{**}$} & \textbf{0.0235$^{**}$} & \textbf{0.0773$^{**}$} & \textbf{0.1068$^{**}$} & \textbf{0.0453$^{**}$} & \textbf{0.0527$^{**}$} & \textbf{0.0317$^{**}$} & \textbf{0.0523$^{**}$} & \textbf{0.0159$^{**}$} & \textbf{0.0211$^{**}$} \\
Improv  & 11.9\%          & 8.50\%          & 15.8\%          & 12.0\%          & 17.5\%          & 16.9\%          & 22.4\%          & 21.1\%          & 13.3\%          & 12.3\%          & 18.9\%          & 17.6\%          & 9.69\%          & 7.17\%          & 11.2\%          & 9.33\%          \\ \hline
FMLP    & 0.0602          & 0.0934          & 0.0311          & 0.0394          & 0.0323          & 0.0524          & 0.0166          & 0.0217          & 0.0676          & 0.0982          & 0.0377          & 0.0447          & 0.0297          & 0.0495          & 0.0143          & 0.0197          \\
DR4SR   & \underline{0.0635$^{**}$} & \underline{0.0993$^{**}$} & \underline{0.0332$^{**}$} & \underline{0.0421} & \underline{0.0345} & \underline{0.0559} & \underline{0.0177$^{**}$} & \underline{0.0230$^{**}$} & \underline{0.0717$^{**}$} & \underline{0.1061$^{**}$} & \underline{0.0400$^{**}$} & \underline{0.0486$^{**}$} & \underline{0.0316$^{**}$} & \underline{0.0524$^{**}$} & \underline{0.0158$^{**}$} & \underline{0.0210$^{**}$} \\
Improv  & 5.48\%          & 6.32\%          & 6.75\%          & 6.85\%          & 6.81\%          & 6.68            & 6.63\%          & 5.99\%          & 6.07\%          & 8.04\%          & 6.10\%          & 8.72\%          & 6.40\%          & 5.86\%          & 10.5\%          & 6.60\%          \\
DR4SR+  & \textbf{0.0687$^{**}$} & \textbf{0.1056$^{**}$} & \textbf{0.0357$^{**}$} & \textbf{0.0449$^{**}$} & \textbf{0.0384$^{**}$} & \textbf{0.0597$^{**}$} & \textbf{0.0198$^{**}$} & \textbf{0.0253$^{**}$} & \textbf{0.0788$^{**}$} & \textbf{0.1136$^{**}$} & \textbf{0.0437$^{**}$} & \textbf{0.0524$^{**}$} & \textbf{0.0353$^{**}$} & \textbf{0.0582$^{**}$} & \textbf{0.0171$^{**}$} & \textbf{0.0231$^{**}$} \\
Improv  & 14.1\%          & 13.1\%          & 14.8\%          & 14.0\%          & 18.9\%          & 13.9\%          & 19.3\%          & 16.6\%          & 16.6\%          & 15.7\%          & 15.9\%          & 17.2\%          & 18.9\%          & 17.6\%          & 19.6\%          & 17.3\%          \\ \hline
GNN     & 0.0570          & 0.0859          & 0.0311          & 0.0384          & 0.0311          & 0.0476          & 0.0167          & 0.0211          & 0.0697          & 0.0958          & 0.0403          & 0.0469          & 0.0242          & 0.0430          & 0.0118          & 0.0166          \\
DR4SR   & \underline{0.0611$^{**}$} & \underline{0.0926$^{**}$} & \underline{0.0324$^{*}$} & \underline{0.0406$^{*}$} & \underline{0.0336$^{**}$} & \underline{0.0525$^{**}$} & \underline{0.0182$^{**}$} & \underline{0.0230$^{**}$} & \underline{0.0736$^{**}$} & \underline{0.1031$^{**}$} & \underline{0.0424$^{**}$} & \underline{0.0498$^{**}$} & \underline{0.0268$^{**}$} & \underline{0.0451$^{*}$} & \underline{0.0129$^{**}$} & \underline{0.0175$^{*}$} \\
Improv  & 7.19\%          & 7.80\%          & 4.18\%          & 5.73\%          & 8.04\%          & 10.3\%          & 8.98\%          & 9.00\%          & 5.60\%          & 7.62\%          & 5.21\%          & 6.18\%          & 10.7\%          & 4.88\%          & 9.32\%          & 5.42\%          \\
DR4SR+  & \textbf{0.0637$^{**}$} & \textbf{0.0953$^{**}$} & \textbf{0.0334$^{**}$} & \textbf{0.0414$^{**}$} & \textbf{0.0351$^{**}$} & \textbf{0.0545$^{**}$} & \textbf{0.0189$^{**}$} & \textbf{0.0238$^{**}$} & \textbf{0.0771$^{**}$} & \textbf{0.1082$^{**}$} & \textbf{0.0442$^{**}$} & \textbf{0.0521$^{**}$} & \textbf{0.0272$^{**}$} & \textbf{0.0471$^{**}$} & \textbf{0.0134$^{**}$} & \textbf{0.0184$^{**}$} \\
Improv  & 11.8\%          & 10.9\%          & 7.40\%          & 7.81\%          & 12.9\%          & 14.5\%          & 13.2\%          & 12.8\%          & 10.6\%          & 12.9\%          & 9.68\%          & 11.1\%          & 12.4\%          & 9.53\%          & 13.6\%          & 10.8\%          \\ \hline
CL4SRec & 0.0653          & 0.0947          & 0.0370          & 0.0441          & 0.0381          & 0.0559          & 0.0215          & 0.0259          & 0.0781          & 0.1075          & 0.0456          & 0.0530          & 0.0322          & 0.0535          & 0.0159          & 0.0212          \\
DR4SR   & \underline{0.0732$^{**}$} & \underline{0.1016$^{**}$} & \underline{0.0423$^{**}$} & \underline{0.0495$^{**}$} & \underline{0.0401$^{**}$} & \underline{0.0600$^{**}$} & \underline{0.0227$^{**}$} & \underline{0.0274$^{**}$} & \underline{0.0821$^{**}$} & \underline{0.1113$^{*}$} & \underline{0.0481$^{**}$} & \underline{0.0551$^{*}$} & \underline{0.0344$^{**}$} & \underline{0.0561$^{**}$} & \underline{0.0174$^{**}$} & \underline{0.0229$^{**}$} \\
Improv  & 12.1\%          & 7.29\%          & 14.3\%          & 12.2\%          & 5.25\%          & 7.33\%          & 5.58\%          & 5.79\%          & 5.12\%          & 3.53\%          & 5.48\%          & 3.96\%          & 6.83\%          & 4.86\%          & 9.43\%          & 8.02\%          \\
DR4SR+  & \textbf{00756$^{**}$}  & \textbf{0.1062$^{**}$} & \textbf{0.0440$^{**}$} & \textbf{0.0517$^{**}$} & \textbf{0.0448$^{**}$} & \textbf{0.0655$^{**}$} & \textbf{0.0247$^{**}$} & \textbf{0.0299$^{**}$} & \textbf{0.0829$^{**}$} & \textbf{0.1140$^{**}$} & \textbf{0.0489$^{**}$} & \textbf{0.0567$^{**}$} & \textbf{0.0363$^{**}$} & \textbf{0.0598$^{**}$} & \textbf{0.0183$^{**}$} & \textbf{0.0241$^{**}$} \\
Improv  & 15.8\%          & 1.12\%          & 18.9\%          & 17.2\%          & 17.6\%          & 17.2\%          & 14.8\%          & 15.4\%          & 6.15\%          & 6.05\%          & 7.24\%          & 6.98\%          & 12.7\%          & 11.8\%          & 15.1\%          & 13.7\%          \\ \hline
\end{tabular}

}
\vspace{-0.2cm}
\end{table*}

%% file: table/ablation.tex
\begin{table}[]
\caption{Abalation study of DR4SR on NDCG@20.}
\label{tab: ablation study}
\vspace{-0.2cm}
\begin{tabular}{l|cccc}
\hline
Dataset        & Beauty          & Sport           & Toys            & Yelp            \\ \hline
SASRec         & 0.0368          & 0.0194          & 0.0448          & 0.0193          \\
DR4SR+         & \textbf{0.0412} & \textbf{0.0235} & \textbf{0.0527} & \textbf{0.0211} \\ \hline
(A) -diversity & 0.0365          & 0.0211          & 0.0470          & 0.0196          \\
(B) pattern    & 0.0181          & 0.0184          & 0.0407          & 0.0141          \\
(C) end-to-end & 0.0026          & 0.0029          & 0.0067          & 0.0035          \\ \hline
\end{tabular}
\vspace{-0.3cm}
\end{table}

%% file: table/runtime.tex
\begin{table}[]
\caption{Time and Space Efficiency Analysis.}
\label{tab: runtime}
\vspace{-0.2cm}
\resizebox{\linewidth}{!}{
\begin{tabular}{c|r|cccc}
\hline
Dataset                & Metric           & Beauty & Sport  & Toys   & Yelp   \\ \hline
\multirow{2}{*}{BASE}  & Runtime(s/epoch) & 7.618  & 15.345 & 13.370 & 17.738 \\
                       & GPU memory (MB)  & 1930   & 2194   & 1968   & 2254   \\ \hline
\multirow{2}{*}{w/ bi-level optimization} & Runtime(s/epoch) & 9.476  & 18.952 & 14.213 & 22.41  \\
                       & GPU memory (MB)  & 2342   & 2626   & 2382   & 2688   \\ \hline
\end{tabular}
}
\end{table}

%% file: content/6.conclusion.tex

In this paper, we have investigated a novel problem concerning the development of training data for sequential recommendations. To tackle this problem, we have introduced a data-centric dataset regeneration framework called DR4SR. Our framework has displayed exceptional cross-architecture generalizability and has underscored the complementary relationship between data-centric and model-centric paradigms. Moreover, we have extended DR4SR to a model-aware version, dubbed DR4SR+, which allows for personalized adaptation of the regenerated datasets to various target models. Additionally, we have conducted in-depth analysis to investigate the potential of the data-centric approach, providing valuable insights. For our future work, we intend to propose a more comprehensive framework that can regenerate various forms of data, such as sequences, graphs, and augmented data. We also plan to explore the integration of large language Models~\cite{huang2024survey, he2023survey} to guide the dataset regeneration process, enabling the generation of data that maintains both collaborative and semantic information.

%% file: content/7.appendix.tex
\input{table/supplement_datasets_results}
\subsection{Pseudo Code of DR4SR and DR4SR+}
The pipeline of DR4SR is detailed in Algorithm \ref{alg: DR4SR}, and the pipeline of DR4SR+ is detailed in Algorithm \ref{alg: DR4SR+}.

\begin{algorithm}[b]
\renewcommand{\algorithmicrequire}{\textbf{Input:}}
\renewcommand{\algorithmicensure}{\textbf{Output:}}
\caption{Pseudo code of DR4SR}
\label{alg: DR4SR}
\begin{algorithmic}[1]
\Require The original sequence dataset $\mathcal{X}$. 
\Ensure A regenerated dataset $\mathcal{X}'$ and scores for each data sample $\mathbf{W} \in \mathbb{R}^{|\mathcal{X}'|}$.

\State Construct a pre-training dataset $\mathcal{X}_{\text{pre}}$ as in Section \ref{Sec: pre-training task}
\While {not converged} \Comment{Pre-training}
    \State Extract a batch of sequence-pattern pairs $(s_u, p_i)$ from $\mathcal{X}_{\text{pre}}$
    \State Get K memories $\{\mathbf{m}'_1, \mathbf{m}'_2, \dots, \mathbf{m}'_K\} = \text{Encoder}(s_u)$ 
    \State Get a category vector $\pi = \text{DiversityPromoter}(p_i)$ 
    \State Get a mixed memory by $\mathbf{m}_{\text{final}} = \sum_{k=1}^{K} \mathbf{\pi}_k \mathbf{m}'_k$
    \State Get a decoded sequence $p_i' = \text{Decoder}(p_i, \mathbf{m}_{\text{final}})$
    \State Optimize the dataset generator with Equation \ref{eq: reconstruction loss}
\EndWhile

\While {not converged} \Comment{Inference}
    \State Extract a sequence $s_u$ from the original dataset $\mathcal{X}$
    \State Get K memories $\{\mathbf{m}'_1, \mathbf{m}'_2, \dots, \mathbf{m}'_K\} = \text{Encoder}(s_u)$
    \State Init the regenerated dataset $\mathcal{X}'$ as an empty set
    \For{$\mathbf{m}'_k$ in $\{\mathbf{m}'_1, \mathbf{m}'_2, \dots, \mathbf{m}'_K\}$}
        \State Init the regenerated pattern $p_i'$ as the <BOS> token
        \While {$v \ne$ <EOS> token}
            \State Get a probability vector $\mathbf{q} = \text{Decoder}(p_i', \mathbf{m}'_k)$
            \If{Generative decoding}
                \State Get item $v \in \mathcal{V}$ with the highest probability
            \Else
                \State Get item $v \in s_u$ with the highest probability
            \EndIf
            \State Update the regenerated pattern $p_i' \leftarrow Concat(p_i', v)$
        \EndWhile
        \State $\mathcal{X}' \leftarrow \mathcal{X}' \cup {p_i'}$
    \EndFor
\EndWhile

\end{algorithmic}
\end{algorithm}

\begin{algorithm}[b]
\renewcommand{\algorithmicrequire}{\textbf{Input:}}
\renewcommand{\algorithmicensure}{\textbf{Output:}}
\caption{Pseudo code of DR4SR+}
\label{alg: DR4SR+}
\begin{algorithmic}[1]
\Require The regenerated sequence dataset $\mathcal{X}'$, a target model $f$ with parameters $\theta$. 
\Ensure A personalizer $g$ with parameters $\phi$ that can assign scores $\mathbf{W} \in \mathbb{R}^{|\mathcal{X}'|}$ for each data sample in $\mathcal{X}'$.

\State Generate a dev dataset $\mathcal{X}'_{\text{dev}}$ by shuffling $\mathcal{X}'$
\While{not converged}
    \State // Lower optimization
    \For{round = 1 $\to T_{\text{lower}}$}
        \State Extract a batch $b$ from $\mathcal{X}'$
        \State Get data sample scores $\mathbf{W} = g(b)$
        \State Optimize $f$ based on Equation \ref{eq: rec loss with weight}
    \EndFor
    \State // Upper optimization
    \State Extract a batch $b_{\text{dev}}$ from $\mathcal{X}'_{\text{dev}}$
    \State Get $\nabla_{\theta} L_{\text{dev}} (\theta^*(\phi))$ with $b_{\text{dev}}$ and Equation \ref{eq: overall}
    \State Get $\nabla_{\phi} \theta^*(\phi)$ by implicit differentiation
    \State Get $\nabla_{\phi} L_{\text{dev}} (\theta^*(\phi))$ as in Equation \ref{eq: implicit gradient}
    \State update $\phi$ with $\nabla_{\phi} L_{\text{dev}} (\theta^*(\phi))$

\EndWhile

\end{algorithmic}
\end{algorithm}

\subsection{Detailed Derivations of Equation \ref{eq: final implicit optimization}}\label{appendix: derivation}
By applying the chain rule, we can compute calculate the implicit gradient of $L_{\text{dev}} (\theta^*(\phi))$ as follows:
\begin{equation}\label{eq: implicit gradient}
    \nabla_{\phi} L_{\text{dev}} (\theta^*(\phi)) = \nabla_{\theta} L_{\text{dev}} (\theta^*(\phi)) \nabla_{\phi} \theta^*(\phi).
\end{equation}
The first term $\nabla_{\theta} L_{\text{dev}} (\theta^*(\phi))$ can be obtained with automatic differentiation (autograd) tools. As for the second term, we notice that $\theta^*(\phi)$ is the optimal point of $L_\text{train}(\theta;\phi)$, so we have:
\begin{equation}\label{eq: minimal point}
    \nabla_{\theta} L_{\text{train}} (\theta^*(\phi), \phi) = 0.
\end{equation}
Then, we calculate the gradient with respect to $\phi$ in Equation \ref{eq: minimal point}:
\begin{equation}
    \nabla_{\theta}^2 L_{\text{train}} (\theta^*(\phi), \phi) \nabla_{\phi} \theta^*(\phi) + \nabla_{\phi} \nabla_{\theta} L_{\text{train}} (\theta^*(\phi), \phi) = 0,
\end{equation}
which leads to:
\begin{equation}
    \nabla_{\phi} \theta^*(\phi) = - (\nabla_{\theta}^2 L_{\text{train}} (\theta^*(\phi), \phi))^{-1} \cdot \nabla_{\phi} \nabla_{\theta} L_{\text{train}} (\theta^*(\phi), \phi),
\end{equation}
where the Hessian inverse $(\nabla_{\theta}^2 L_{\text{train}} (\theta^*(\phi), \phi))^{-1}$ can be approximated to $\sum_{n=0}^K (I - \nabla_{\theta}^2 L_{\text{train}} (\theta^*(\phi))^n$ with K-truncated Neumann series\cite{aux_learn1, aux_learn2}. Therefore, we can obtain the following results:
\begin{equation}
    \nabla_{\phi} \theta^*(\phi) = - \sum_{n=0}^K (I - \nabla_{\theta}^2 L_{\text{train}} (\theta^*(\phi))^n \cdot \nabla_{\phi} \nabla_{\theta} L_{\text{train}} (\theta^*(\phi), \phi).
\end{equation}
Finally, the implicit gradient of $L_{\text{dev}} (\theta^*(\phi))$ is:
\begin{equation}
    \nabla_{\phi} L_{\text{dev}} = - \nabla_{\theta} L_{\text{dev}} \cdot \sum_{n=0}^K (I - \nabla_{\theta}^2 L_{\text{train}})^n \cdot \nabla_{\phi} \nabla_{\theta} L_{\text{train}},
\end{equation}
where the parameters are omitted for the sake of brevity.

\subsection{Computational Complexity}\label{sec: computational_complexity}
The scale of datasets can be measured with the number of sequences N, maximum sequence length L, maximum pattern length M, and the number of extracted patterns q (empirically, q is of the same order of magnitude as N). We denote the diversity factor as K.

The pre-training dataset is constructed with sequential pattern mining, which is implemented with Seq2Pat~\cite{Seq2Pat} and its complexity is $O(NMq)$. Empirically, this process can be accelerated with offline pre-computation or batching technique~\cite{Seq2Pat}. During pre-training, the regenerator needs to process each sequence. Considering the quadratic complexity of the self-attention mechanism, the overall complexity is $O(qL^2d)$. During inference, the complexity is $O(KNL^2d)$, as we need to regenerate K patterns from each sequence. Meanwhile, autoregressive decoding can be further accelerated with efficient decoding~\cite{efficient_llm_inference}. As for dataset personalization, denote the complexity of the target model as $a$. The additional overhead incurred by the bi-level optimization stems from the upper optimization conducted every $T_{\text{lower}}$ rounds. In the upper optimization, complexity mainly results from the implicit gradient computation, which necessitates (K+2) backward for vector-Jacobi product\cite{aux_learn2}. Herein, K represents the truncated number of Neumann series, which is fixed at 3. Consequently, the complexity merely amounts to $O\left(\left(1 + \frac{5}{T_{\text{lower}}}\right)a\right)$. Therefore, with some acceleration strategies, DR4SR and DR4SR+ is scalable when applied to large datasets.

\subsection{Detailed Hyper-parameters}\label{appendix: hyper-parameter}
We first introduce the hyper-parameter settings of target models: For GRU4Rec, we search the layer number among [2, 3], and dropout probability among [0.2, 0.5]. For SASRec, the layer number is 2, the dimension of hidden layers is 128, and the dropout probability is 0.5. For FMLP, the layer number is 2, the dimension of hidden layers is 256, and the dropout probability is 0.5. For GCE-GNN, we first replace its graph encoder with LightGCN\cite{LightGCN} and its sequential encoder with SASRec\cite{SASRec}, which can significantly improve its performance. Then, we search the number of graph convolution layers among [2, 3]. For CL4SRec, we set the weight of contrastive learning loss to 0.1, the crop/mask/reorder rate in data augmentation is set to 0.2/0.7/0.2, and other hyper-parameters are the same as SASRec.

Then we consider particular hyper-parameters of DR4SR. For dataset regeneration, we set sliding window size $\alpha$ to 5 for Beauty and 10 for others, and we set the threshold $\beta$ to 2 for all datasets. We fix the diversity number $K$ to 5, and the generative decoding probability $\gamma$ to 0.1. For DR4SR+, we set $T_{\text{lower}}$ to 30, and adopt SGD as the upper optimizer with learning rate and weight decay searched among [1e-3, 1e-2, 1e-1].

\subsection{Generalization to More Datasets}

We provide the results on two additional datasets Amazon-book and Gowalla in Table \ref{tab: additional_overall_performance}, where we can find that the proposed method can be generalized to datasets with large-scale or different distribution.

%% file: table/supplement_datasets_results.tex
\begin{table*}
\caption{The overall performance on two additional datasets.}
\label{tab: additional_overall_performance}
\vspace{-0.2cm}
\resizebox{\textwidth}{!}{
\begin{tabular}{l|l|ccc|ccc|ccc|ccc|ccc}
\hline
        & Metric & GRU4Rec & DR4SR           & DR4SR+          & SASRec & DR4SR           & DR4SR+          & FMLP   & DR4SR           & DR4SR+          & GNN    & DR4SR           & DR4SR+          & CL4SRec & DR4SR           & DR4SR+          \\ \hline
Gowalla & R@20   & 0.1297  & \underline{0.1325} & \textbf{0.1392} & 0.1672 & \underline{0.1786} & \textbf{0.1856} & 0.1710 & \underline{0.1811} & \textbf{0.1833} & 0.1274 & \underline{0.1331} & \textbf{0.1348} & 0.2012  & \underline{0.2064} & \textbf{0.2095} \\
        & N@20   & 0.0582  & \underline{0.0615} & \textbf{0.0634} & 0.0833 & \underline{0.0912} & \textbf{0.0934} & 0.0851 & \underline{0.0903} & \textbf{0.0922} & 0.0627 & \underline{0.0653} & \textbf{0.0662} & 0.0987  & \underline{0.1032} & \textbf{0.1048} \\ \hline
Book    & R@20   & 0.0223  & \underline{0.0238} & \textbf{0.0245} & 0.0308 & \underline{0.0343} & \textbf{0.0372} & 0.0388 & \underline{0.0432} & \textbf{0.0444} & 0.0301 & \underline{0.0332} & \textbf{0.0346} & 0.0367  & \underline{0.0378} & \textbf{0.0386} \\
        & N@20   & 0.0087  & \underline{0.0098} & \textbf{0.0112} & 0.0125 & \underline{0.0140} & \textbf{0.0154} & 0.0162 & \underline{0.0181} & \textbf{0.0188} & 0.0129 & \underline{0.0141} & \textbf{0.0148} & 0.0153  & \underline{0.0161} & \textbf{0.0165} \\ \hline
\end{tabular}}
\end{table*}